# MOSAIC: A Multi-Objective Optimization Framework for Sustainable Datacenter Management

Sirui Qi
*Colorado State University*
Fort Collins, USA
alex.qi@colostate.edu

Dejan Milojicic, Cullen Bash
*Hewlett Packard Labs*
Milpitas, USA
{dejan.milojicic, cullen.bash}@hpe.com

Sudeep Pasricha
*Colorado State University*
Fort Collins, USA
sudeep@colostate.edu

*Abstract*— In recent years, cloud service providers have been building and hosting datacenters across multiple geographical locations to provide robust services. However, the geographical distribution of datacenters introduces growing pressure to both local and global environments, particularly when it comes to water usage and carbon emissions. Unfortunately, efforts to reduce the environmental impact of such datacenters often lead to an increase in the cost of datacenter operations. To co-optimize the energy cost, carbon emissions, and water footprint of datacenter operation from a global perspective, we propose a novel framework for m̲ulti-o̲bjective sus̲ta̲inable dat̲ac̲enter management (MOSAIC) that integrates adaptive local search with a collaborative decomposition-based evolutionary algorithm to intelligently manage geographical workload distribution and datacenter operations. Our framework sustainably allocates workloads to datacenters while taking into account multiple geography- and time-based factors including renewable energy sources, variable energy costs, power usage efficiency, carbon factors, and water intensity in energy. Our experimental results show that, compared to the best-known prior work frameworks, MOSAIC can achieve 27.45× speedup and 1.53× improvement in Pareto Hypervolume while reducing the carbon footprint by up to 1.33×, water footprint by up to 3.09×, and energy costs by up to 1.40×. In the simultaneous three-objective co-optimization scenario, MOSAIC achieves a cumulative improvement across all objectives (carbon, water, cost) of up to 4.61× compared to the state-of-the-arts.

## I. Introduction

In the last decade, more and more applications have been deployed to cloud datacenters, such as scientific computing, image processing, anomaly detection, recommendation engines, etc. [1]. Today, cloud datacenters hosted by providers such as Amazon and Google are built and operated across multiple geographical sites to provide robust cloud services to users. The geographical distribution of datacenters has many advantages, including improved performance and lower costs by bringing datacenters closer to consumers, and better resilience to catastrophic datacenter failures (e.g., due to hurricanes and other extreme weather events) by distributing workloads to different geographical locations.

However, datacenters across all scales (small, medium, large) have a significant energy consumption and also create pressure on local and global environments. Today, 2% of electricity use in the United States and 1% of worldwide electricity usage can be attributed to datacenters [2]. Further, 0.6% of global greenhouse gas emissions come from datacenters [3], which is the same as the entire worldwide airline industry [4]. Datacenters also consume large quantities of water for cooling, e.g., a large datacenter can use up anywhere between 1 million and 5 million gallons of water a day — as much as a town of 10,000 to 50,000 people [5], which creates a major burden on local water supplies.

With global climate change becoming more of a reality every day, many countries are starting to tighten their environmental policies, which is forcing cloud service providers to recognize the need for realizing sustainable datacenters that have a lower environmental (water, carbon emission) footprint [6], [7]. Therefore today, cloud service providers must not only focus on minimizing their operating costs (to maximize profits), but also consider the environmental impacts of their datacenters. From the perspective of managing datacenter energy (electricity use) costs, cloud workloads should be assigned to datacenter locations where there are inexpensive energy sources. However, from the perspective of improving sustainability, it is better to migrate workloads to locations that can provide cleaner (e.g., solar or wind) energy sources. These two perspectives are vital to datacenter management but are usually in conflict. For instance, electricity costs may be the lowest at a datacenter location that uses coal (brown energy) for electricity generation, whereas a datacenter location with sufficient wind power may have high (green energy) costs. Hence, mapping cloud workloads to datacenters so that energy cost and sustainability goals can be met simultaneously is a challenging problem.

The geographical distribution of datacenters provides many new opportunities for cloud service providers to intelligently manage datacenters, beyond decisions related to green and brown energy use. For instance, many locations have time-of-use (TOU) electricity pricing [8] to encourage moving datacenter workloads to off-peak periods (e.g., during nighttime), where electricity prices might be 10× lower than in the peak period. Datacenters at many locations may also be equipped with green energy based cooling techniques such as free air cooling [9] and thermosyphons [10], which do not require coolant pumping or mechanical refrigeration in datacenters. Such mechanisms have compelling benefits, e.g., free air cooling can replace mechanical refrigeration components (e.g., cooling room air conditioners) with direct air exchangers to reduce both environmental and energy overheads of datacenters [9], [11]. However, free air cooling has more stringent requirements for outdoor temperature and dew point according to ASHRAE [12]. Thus, to jointly optimize energy costs and sustainability goals of datacenters, cloud service providers must consider both time- and geographical-based factors.

Prior works on energy-aware datacenter workload management have usually formulated a single objective optimization problem that focuses on minimizing either datacenter energy cost [13] or an isolated environmental impacts such as carbon emissions or water usage (e.g., [14], [15]). However, today we must consider diverse sustainability goals, and, as discussed above, these are usually in conflict with reducing energy costs. Hence, cloud service providers must jointly optimize energy cost and environmental (water-use and carbon emission) impacts. To address this new challenge in datacenter management, we propose a novel framework for multi-objective sustainable datacenter management (MOSAIC) to co-optimize energy costs, carbon emissions, and water footprint of datacenters. Our proposed framework performs smart design space exploration to generate Pareto-optimal solutions that co-optimize the energy costs and environmental impacts of geographically distributed datacenter operations. Our novel contributions can be summarized as follows:

- We comprehensively model the energy profile, carbon emissions, and water usage of datacenters while capturing their time- and geography-based differences.
- We propose a new multi-objective optimization framework that combines self-guided local search and collaborative decomposition-based evolutionary algorithm to optimize the three objectives (energy costs, carbon emissions, water use) simultaneously for datacenters.
- We compare our framework with the state-of-the-art datacenter management frameworks and find that MOSAIC outperforms other frameworks in Pareto Hypervolume, convergence speed, scalability, and cumulative solution quality.

The rest of this paper is organized as follows. In Section II, we review relevant prior works in both single objective and multi-objective datacenter management. We describe our models in Section III. Sections IV and V outline our problem formulation and proposed framework for sustainable datacenter management. The comparison methodology and experiment results are presented in Section VI. Finally, we present concluding remarks in Section VII.

## II. RELATED WORK

Resource allocation and management in the cloud has been studied for many years [16]. Researchers have recognized objectives such as quality of service [17], cost [18], [19], [20], performance [21], revenue [22], fault rate [23] and resource utilization rate [24], for which single-objective optimization techniques have been proposed.

Long et al. [25] proposed a game theory-based approach to solve a task assignment problem in collaborative edge and cloud environments. The game theory is formulated into a non-cooperative game among multi-agents (multiple edge datacenters) to minimize energy costs of task scheduling. The game theory-based approach shows improved performance and speedup for such problems by reaching Nash equilibrium, when compared to state-of-art frameworks.

Yang et al. [26] proposed a deep reinforcement learning enhanced greedy optimization algorithm for two-stage task sequencing and task allocation, to maximize a system's gain which is defined as the value of completed tasks minus system operation costs. The authors first deployed a deep reinforcement learning module to predict the best allocation sequence for each arriving batch of tasks. Following the best allocation sequence, the authors proposed a greedy strategy that allocates tasks to datacenter servers one by one, in an online setting, to maximize the total gain increase.

Hogade et al. [13] proposed the genetic algorithm load distribution (GALD) approach to optimize energy costs in datacenters. Techniques such as TOU pricing, peak shaving, net metering, and utilizing on-site clean energy sources were all considered in their framework to fully reduce the energy costs, but these also created a complex design space. Genetic algorithms (GAs) that were used in the GALD framework are adaptive heuristic search algorithms inspired by natural selection and have been shown to be efficient in complex design space exploration including workload scheduling. GAs have also been proven effective in multi-objective problems [27], but are faced with slow convergence rates in many multi-objective problems [28].

Besides single objective optimization approaches such as the ones described above, multi-objective datacenter resource management has also been receiving attention in recent years. Various multi-objective optimization approaches have been proposed using techniques such as simulated annealing (SA), and non-dominated sorting [29], [30].

Liu et al. [29] proposed a holistic optimization framework for mobile cloud workload computing. A triple-objective optimization (TOO) problem is formulated in their mobile cloud management scenario, which involves optimizing energy consumption, system reliability, and quality of service. SA was utilized in their framework to generate a Pareto optimal solution set and was shown to be effective in providing trade-offs across the three objectives. SA is a probabilistic approach to approximate the global optimum by mimicking the slow cooling of metal but it often experiences performance degradation and slowdown in larger design space exploration problems compared to GAs.

Bi et al. [30] proposed a decomposition-based multi-objective evolutionary algorithm with Gaussian mutation and crowding distance (DMGC), to jointly optimize energy cost and revenue of workload scheduling in datacenters. Compared with GA, the authors were able to converge faster during design space exploration and provide a better Pareto optimal solution set. Their speedup came from the Gaussian mutation in which random offspring generation follows the Gaussian distribution pattern, taking parent solutions as the center of the distribution. In this manner, DMGC was able to explore the design space faster than GAs. The quality improvement with their Pareto optimal solution set is due to the fact that DMGC was able to preserve more diverse designs in their solution set through a crowding distance mechanism. Crowding distance measures a solution's degree of alienation compared to its neighboring solutions. A solution set with higher crowding distance helps the framework maintain a diverse solution set where solutions are farther apart from each other in the design space. Such a diverse solution set provides diverse parents for the later crossover operator and hence prevents DMGC from falling into local optima. Aided by the diverse solution set, DMGC was able to more comprehensively explore the design space and generate a better Pareto optimal set. However, our analysis indicates that the crowding distance approach may experience performance degradation because such an approach may overvalue the crowding distance while ignoring high quality designs. Low quality designs with higher crowding distance can eliminate high quality designs with lower crowding distance in DMGC. In a large design space, crossover between high quality designs with relatively low crowding distances will likely generate better offspring, which DMGC often fails to guarantee.

At the same time, modeling key constraints and performance factors are vital for datacenter management, for single-objective optimization as well as multi-objective optimization, while considering deadlines, thermal constraints, co-location, etc. In [31], a deadline constrained task arriving model and utility gain function associated with deadline overhead are introduced to manage an oversubscribed heterogeneous datacenter. In [32], [33], the authors predicted the thermal implications of task allocation on different computing components, and utilized thermal prediction models to avoid thermal constraint violations in datacenters. In [34], [35], the authors studied execution time degradation of tasks under co-location interference effects due to computing and memory resource sharing. By combining both thermal modeling and co-location interference modeling, a novel framework was developed in [36] to maximize utilization of servers in datacenter while easing hotspot phenomena.

In our work, we have investigated drawbacks of state-of-the-art frameworks [13], [29], [30] in datacenter management such as slow convergence speed and lack of scalability for a large design space. To overcome these shortcomings, we propose a novel multi-objective optimization framework called MOSAIC which combines local search and evolutionary algorithm for datacenter management. To the best of our knowledge, MOSAIC is the first framework to co-optimize energy costs, carbon emissions, and water use in datacenters.

## III. SYSTEM MODEL

In our framework, workloads originate from different locations and are propagated to a 2-tier distribution system. In the first tier, a global manager assigns workloads from their origin to different datacenters. In the second tier, a local scheduler at each datacenter location assigns workloads at the datacenter to different nodes inside the datacenter.

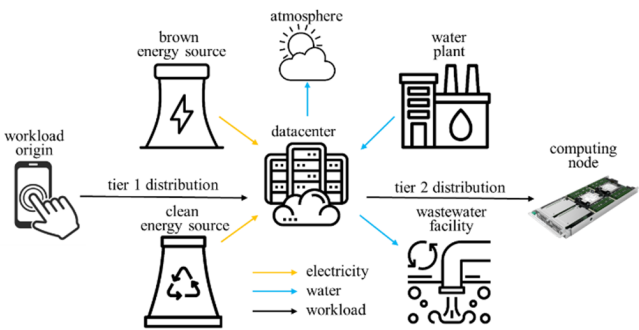

Fig. 1. Electricity/water/workload flow in a 2-tier distribution. A workload is assigned to a datacenter from its origin during tier 1 distribution. After arriving at a datacenter, a workload is assigned to a computing node during tier 2 distribution.

Fig. 1 illustrates such a 2-tier distribution system. To co-optimize the energy cost and environmental impacts of global datacenter operation, our framework not only controls the workload distributions across and within datacenters but also the corresponding energy payment plans (e.g., for annual clean contracts, discussed in Section III.*D*). Each datacenter is modeled in terms of its power use, carbon emissions, water use, energy costs, and workload, as discussed in the rest of this section.

### A. Power Model
#### 1) Datacenter Layout

Each datacenter has N computing nodes installed in racks [37] and the racks are arranged in a standard hot-aisle/cold-aisle configuration [38]. Computing nodes have different energy profiles and performance according to their types, e.g., a 4-core server node and a 12-core server node will differ in their throughput/latency and energy consumption. Cooling room air conditioning (CRAC) units are installed in datacenter rooms to cool down computing nodes. Besides mechanical cooling systems, free air cooling systems are installed in a subset of datacenters and utilized when outdoor temperature and dew point both allow for it. Free air cooling systems pump cold outdoor air directly into the datacenter through direct air exchangers. When the outdoor temperature and dew point are in feasible regions, the datacenter will switch to free air cooling from mechanical cooling [9]. We assume that high performance filters [39] are installed in direct air exchangers to minimize risks from air pollution.

#### 2) Datacenter Power Components

The power consumption of a datacenter $d$ contains three components in our model: information technology (IT) load $P_{IT,d}$, cooling $P_{Cooling,d}$, and Internal Power Conditioning System (IPCS) $P_{IPCS,d}$ [40].

IT load represents the total power consumption of computing nodes in datacenters and has a direct correlation on with executed workloads. Consider $L$ different workload types executed by $A$ active server nodes. We assume a fixed P-state for each node, which determines the voltage and frequency of the node during execution. The active power $AP$ of a node depends on workload type $i$ and the type of node $j$. $IP$ is idle power of any node that is not active. Then the total power of the IT load at a datacenter can be calculated as:

$$P_{IT} = \sum_{i}^{L}\sum_{j}^{A} AP_{ij} + \sum_{k}^{N-A} IP_k \quad (1)$$

Cooling power $P_{Cooling}$ has three components: CRAC units, chillers, and supporting equipment such as cooling tower, fans, etc. The power consumption of chillers and the supporting equipment is estimated to be approximately equal to the power consumption of CRAC units in [11]. Hence, we can estimate the total cooling power at a datacenter as:

$$P_{Cooling} = P_{CRAC} + P_{Chiller} + P_{Support} = 3 \times P_{CRAC} \quad (2)$$

To estimate the power consumption of CRAC units, we use the coefficient of performance ($CoP$), which is defined in [11] as the ratio of removed heat to the amount of necessary power to remove such heat. The CRAC power consumption thus can be estimated as:

$$P_{CRAC} = P_{IT} / CoP \quad (3)$$

The IPCS comprises power management components such as server-to-server connections, power supply units, power distribution units, and uninterrupted power supply units. The total power consumption of IPCS correlates to the IT load and can be estimated as [40]:

$$P_{IPCS} = 0.13 \times P_{IT} \quad (4)$$

The total power consumption across all D datacenters owned by a cloud service provider can then be calculated at any instance by adding the power consumed at each datacenter.

### B. Water Model

To comprehensively model the water usage of datacenters, we consider both site-based (from cooling) and source-based (from electricity generation) water consumption. As shown in Eq. (5) below, $V_{E,i}$, $V_{B,i}$, and $V_{S,i}$ denote volumes of evaporative, blowdown-to-wastewater-facility, and source water consumption of datacenter $i$ respectively. By adding together $V_{E,i}$, $V_{B,i}$, and $V_{S,i}$ we can estimate the global water usage across all D datacenters ($V_{All}$):

$$V_{All} = \sum_i^D (V_{E,i} + V_{B,i} + V_{S,i}) \quad (5)$$

Direct water consumption is common in mechanical cooling datacenters, where the water is an expendable coolant. Water plants provide potable water to datacenters, while the wastewater treatment facility is responsible for purifying industrial wastewater from the datacenter [41]. We estimate the volume of direct water consumption through all means of water outflows, which are evaporative water through the cooling tower and blowdown water to the wastewater facility.

Evaporative water consumption $V_E$ comes from water evaporation through the water-cooling tower and can be calculated as:

$$V_E = E_{IT}/H_{Water} \quad (6)$$

where $E_{IT}$ represents the heat generated by the IT infrastructure and $H_{Water}$ is the latent heat of the water.

The volume of blowdown water $V_B$ to the wastewater treatment facilities is another part of the site-based water outflow. We assume that datacenters cycle potable water until the concentration of dissolved solids is roughly $C$ times the supplied water [42]. Hence, we estimate the volume of blowdown water as:

$$V_B = V_E/(C - 1) \quad (7)$$

Lastly, source-based water consumption $V_S$ primarily comes from the electricity generation process when brown energy sources such as coal are in use. Modern power grids usually utilize different energy sources and their brown energy ratios vary across locations. The water intensity in electricity generation is represent by the energy water intensity factor ($EWIF$) metric and its value depends on the local energy infrastructure. For instance, $EWIF$ in Florida is 0.53 L/kWh while in Texas it is 1.67 L/kWh [43], i.e., 1.67 Liters of water are used when generating 1 kWh of electricity in Texas. We calculate source-based water consumption based on the datacenter's energy consumption $E$ (computing, cooling, and switching) as:

$$V_S = E \times EWIF \quad (8)$$

### C. Carbon Model

Carbon dioxide has become the biggest source of greenhouse gas emissions [44]. To reduce carbon emissions in datacenters, prior efforts consider electricity-based carbon emissions but usually ignore water-use-based carbon emissions. Many datacenters consume a large amount of potable water and potable water generation involves industrial treatment which creates extra carbon emissions. One of our novel contributions in this work is to find correlations over carbon, water, and energy use in datacenters. We thus analyze the carbon footprint from not just electricity generation but also potable water usage and wastewater treatment.

Our analysis reveals that mechanical cooling systems and datacenter locations may introduce a larger extra carbon footprint than expected. The overall carbon emission $M_{All}$ of D datacenters can be calculated using Eq. (9) below, where $M_{Electricity,i}$ and $M_{Water,i}$ are the mass of electricity-based and water-based carbon emitted at datacenter $i$ respectively:

$$M_{All} = \sum_i^D (M_{Electricity,i} + M_{Water,i}) \quad (9)$$

The reason why we consider geographical difference $i$ in electricity-based carbon emission $M_{Electricity,i}$ is that carbon intensity in electricity generation is highly dependent on local energy infrastructures. For instance, a location may offer cheap electricity but heavily rely on carbon-intensive energy sources such as coal. Meanwhile, another location may have a higher electricity price but primarily rely on zero-carbon energy sources such as wind power. Carbon Factor $CF$ is a metric that measures the mass of emitted carbon during the process of electricity generation at different locations. From the geographical $CF$ in [45], we formulate Eq. (10) to estimate the mass of electricity-based carbon emissions $M_{Electricity}$ due to the amount of brown energy $E_B$ used in a datacenter:

$$M_{Electricity} = E_B/CF \quad (10)$$

Potable water production and wastewater treatment also contribute to a datacenter's carbon emissions, as discussed earlier. We assume water plants and wastewater treatment facilities use electricity from the local power grid. The potable usage is the sum of blowdown water $V_B$ and evaporative water $V_E$. The amount of wastewater generation is the blowdown water $V_B$ from a datacenter to the wastewater facility. Considering the energy intensities for potable water production $I_P$ and wastewater treatment $I_W$ (representing the energy consumption per unit of water treatment), we estimate the mass of water-based carbon emissions $M_{Water}$ as:

$$M_{Water} = [(V_B + V_E) \times I_P + V_B \times I_W]/CF \quad (11)$$

### D. Energy Cost Model

We consider three price models that are relevant to estimating the energy costs associated with distributing workloads to datacenters at different locations: (a) TOU pricing, (b) clean premium, and (c) annual clean contract.

Time-of-use (TOU) pricing is vital for our framework to determine efficient workload distributions that can avoid peak periods (of electricity pricing) at each location. We consider TOU pricing from electricity providers at each datacenter location. However, solely considering TOU pricing can result in unexpected overheads in environmental impacts. This is because a cheap energy source may not always be a green energy source. For instance, off-peak periods usually occur at midnight when there is no solar energy available.

To enable comprehensive and realistic trade-offs between energy cost and environmental impacts, we consider two additional factors: clean premium and annual clean contract.

Power companies at many locations offer clean premium as an extra charge model for users to purchase clean energy from the power grid [46]. Once users pay the extra clean premium over the original electricity price, the electricity provided to the user from the power grid will include clean energy sources instead of brown or mixed (green and brown) energy sources. Besides promoting sustainability, the clean premium approach enables better configurability for cloud service providers. For instance, consider a datacenter that consumes two units of energy at a local power market. Cloud service providers can choose to pay a clean premium on one unit of energy and therefore get one guaranteed unit of clean energy. Meanwhile, the other unit of energy consumed by the datacenter can be considered as being based on brown or mixed energy. By selectively paying for clean premiums, cloud service providers can balance energy costs and environmental impacts. Local power markets in places such as San Francisco and Denver already make clean premiums available to users. We consider such location-specific clean premium costs for electricity usage at a datacenter in our work.

Compared with clean premium, an annual clean contract is a cheaper way for datacenters to access clean energy but so far it only exists in the Texas area of the United States. Several power providers such as Gexa Energy in Texas provide 24-hour all-year-round electricity from green (renewable) sources. However, the annual clean contract still offers an electricity price that is higher than off-peak TOU price. Further, in order to sign an unchangeable annual clean contract, annual energy use of a datacenter needs to be estimated in advance. We consider such an annual clean contract for datacenters in the Texas area.

Our MOSAIC framework has the ability to judiciously balance TOU prices, clean premiums, and annual clean contracts when distributing workloads across datacenters at different locations. The geography- and time-based differences in TOU pricing and annual clean contracts influence our framework's distribution plans on workloads. The design space not only becomes complex due to TOU pricing and annual clean contracts but also is expanded because of the need to self-decide the amount of clean premiums at every datacenter location to co-optimize the objectives of energy cost, water use, and carbon emissions.

*E. Workload Model*

We consider a rate-based workload model, where the cloud workload arrival rate can be estimated over a decision interval called an epoch [47]. The epoch is assumed to be one hour in our work, and 24 epochs equals a full day. Workload arrival rates can be reasonably approximated as constant in an epoch for cloud datacenters today [48]. As shown in Eq. (12), our framework maps the global arrival rate $GAR_i$ of workload $j$ into local arrival rates $AR_{i,j}$ across the $D$ datacenters:

$$GAR_j = \sum_i^D AR_{i,j} \quad (12)$$

The following sections present our problem formulation and describe our MOSAIC framework in detail.

IV. PROBLEM FORMULATION AND ASSUMPTIONS

We consider a cloud service provider managing datacenters at different locations across the United States. There are a large number of datacenters outside of the United States, but for this work, we only consider datacenters inside the United States to maintain source consistency in geographical factors such as *EWIF*. Besides, the United States covers a relatively large area spanning multiple time zones. Therefore, datacenters across the United States have diverse time- and geography-based factors, creating a complex datacenter management problem.

Within the United States, our framework maps the workloads coming in from various locations to datacenters and then to specific computing nodes insider the datacenters. In each epoch (hour), a global workload distribution plan should contain two parts: *(i)* workloads to be distributed to each location, and *(ii)* an estimate of the amount of clean premium at each location. Such a distribution plan is expected to simultaneously co-optimize three objectives for sustainable datacenter management: *(i)* energy cost, *(ii)* carbon emissions, and *(iii)* water usage.

We assume that datacenters are under-subscribed and that they have enough computational resources to finish any workloads assigned to them before deadlines. This is a realistic assumption as cloud service providers today typically overprovision datacenters to meet peak demand scenarios. The global datacenter management approach is described in the next section (Section V). Once workloads have been assigned to each datacenter in an epoch, we use a local workload scheduling policy (shared by all datacenters) to schedule workloads to compute nodes in the datacenter. This policy primarily depends on workload type and node type and derives from the list scheduling approach, which in turn relies on an ordered list of available heterogeneous compute nodes in a datacenter to guide the mapping [49].

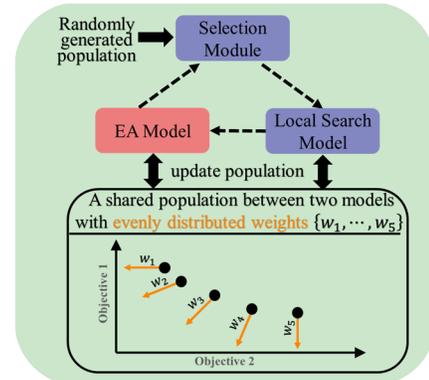

Fig. 2 MOSAIC framework overview

V. MOSAIC FRAMEWORK

Our proposed MOSAIC framework for global workload distribution utilizes a novel hybrid search approach that combines self-guided local search and a collaborative decomposition-based Evolutionary Algorithm (EA), as shown in Fig. 2. Through the same objective decomposition methodology, local search and EA in our framework share their knowledge during design space exploration and jointly optimize the multi-objective datacenter management problem.

As illustrated in Fig. 2, a randomly generated population is input to a *Selection Module* for local search starting point filtering. The *Selection Module* randomly picks starting points in the beginning due to lack of knowledge of local search history. These starting points are locally searched in their neighboring design space by the *Local Search Model* based on improvement in their weighted sums. The weighted sum calculation is further discussed in Section III.*B*, which transforms multiple objective comparison to single objective comparison. After that, the resulting local search endpoints replace original starting points in the population but inherit the same weight vectors for further weighted sum calculation. The update frequency during local search of each starting point is recorded in an update table, which is utilized by the *Selection Module* to decide future local search starting points. After local search, our *EA Model* further explores the design space and helps the *Local Search Model* jump out of local optima in a collaborative way. Hence, the population is further updated by the *EA Model* based on the same set of weight vectors used in local search. This process iteratively updates the population with repeated invocations (each iteration results in a new "generation" of population that have evolved from a previous iteration's "generation"), till a user-defined termination criteria is reached.

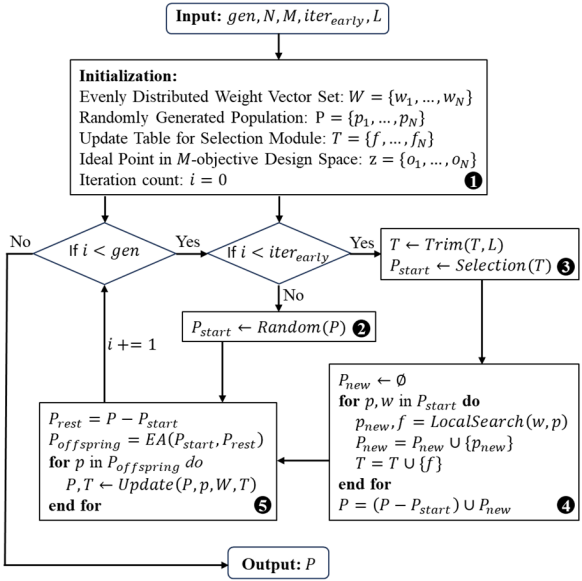

Fig. 3 MOSAIC multi-objective optimization algorithmic flow

Fig. 3 summarizes the algorithmic flow for our framework, which includes self-guided local search (blocks 2-4) and collaborative decomposition-based EA (block 5). The input includes the maximum number of generations to consider for iterative optimization $gen$, population size $N$, number of objectives $M$, number of early iterations for random local search $iter_{early}$, and maximal length $L$ for the update table. Based on the inputs, the initialization process (block 1) creates an evenly distributed weight vector set $W$ for our weighted sum objective decomposition approach (discussed in Section V.*B*), a randomly generated population $P$ to start the optimization, an update table $T$ for local search starting point selection, and an ideal point $z$ used during objective decomposition. The output consists of $N$ Pareto optimal designs for $M$ objectives. The objective values of each design $p$ are calculated based on models introduced in Section III. Details of self-guided local search and collaborative decomposition-based EA in MOSAIC are discussed in the following subsections.

*A. Self-Guided Local Search*

Local search is commonly used in design space exploration where knowledge of the true Pareto front is lacking. However, guiding the direction of local search is challenging for many reasons. An aimless local search model will face performance and speed overheads during exploration in large design spaces (such as in our problem). Hence, a local search history update table is embedded into the starting point selection module to select starting design (solution) points for effective local search that improves the quality of solutions.

The update table records each design point's update frequency during local search. The update frequency is calculated by dividing the update time with local search time and has a value between 0 and 1. A higher update frequency $f_i$ of a starting design point $i$ indicates that the local search model found several good design points in $i$'s neighboring space in the past. Thus, it may be easier for the local search model to find a new good design point in $i$'s neighborhood in the design space during the next round of local search.

The local search model starts with a random local search (block 2) to create an update table $T$ for subsequent access by the selection module $Selection(T)$. The selection module picks a starting design point set $P_{start}$ with higher average update frequency than others, which includes the starting design points with the most potential for effective local search. Additionally, only the previous $L$ time's update frequency is preserved in the update table to calculate the corresponding average update frequency for a design point. Older update frequency data is trimmed from the update table (block 3). This is done to prioritize data only from recent local searches, which in our experience is a more effective strategy than considering data over a lengthy duration of exploration.

At each generation, the local search function $LocalSearch(w, p)$ evaluates each design point $p$ from $P_{start}$ and its corresponding weight vector $w$ (block 4). An empty set $P_{new}$ is created to store all endpoints $p_{new}$ from the local search function. The update table $T$ then records the update frequency $f$ of each design point. Lastly, all local search design endpoints replace their corresponding starting points in $P$ to create an enhanced population that improves the outcomes from our EA approach, which is discussed next.

*B. Collaborative Decomposition-Based EA*

Evolutionary algorithms (EAs) have the desirable ability to jump out of local optima, but they can suffer from slow convergence speed due to biological evolution mechanisms used in the algorithms. We decided to utilize EA to help our local search model to jump out of its local optima during design space exploration with our multi-objective datacenter management problem. In recent years, decomposition-based EAs such as [50] have shown promising results. Existing decomposition-based EAs usually utilize the Tchebycheff approach [51] to decompose multiple objectives into a single artificial objective

$g$ using a set of $N$ uniformly spread weight vectors $W = \{w_1, ..., w_N\}$ in the following manner:

$$g(x|w, z) = \max_{1 \leq i \leq M}\{w_i|Obj_i(x) - z_i|\} \quad (13)$$

From Eq. (13), we know that artificial objective $g$ will always be the largest weighted objective and therefore a multi-objective problem is decomposed into a single objective problem. However, the Tchebycheff approach is a coarse decomposition method that only optimizes a single objective and ignores plenty of potential optimization directions in the design space. Hence, we do not use this approach. Instead, a weighted sum approach is used as the decomposition method in our EA model to provide fine-grained optimization directions in design space. As shown in Eq. (14), the weighted sum of all objectives is considered an artificial objective for optimization in our approach:

$$g(x|w, z) = \Sigma_{i=1}^{M}\{w_i|Obj_i(x) - z_i|\} \quad (14)$$

Our EA model realizes knowledge propagation by purposefully distinguishing locally-searched points $P_{start}$ and non-locally-searched points $P_{rest}$. The crossover in our EA model always occurs between $P_{start}$ and $P_{rest}$ so that local search knowledge is propagated to locally-searched points (block 5). Such crossover can produce more high quality design points and update non-locally-searched design points. Subsequently, our model mutates the crossover offspring to further explore the design space. Thus, our EA model avoids performing a local search on the whole population while still improving the overall population quality by expanding the exploration space. The generated offspring are used to update the population via the function $Update(P, p, W, T)$, where each design point $p$ in offspring $P_{offspring}$ is randomly compared with design points in $P$ using the weighted sum function (Eq. (14)). In addition to updating the population, $Update(P, p, W, T)$ also renew the update table $T$. This is because a new neighboring space emerges once previous design points are replaced by new design points from the EA model. Hence, the previous entry in the update table for this design point is not valid anymore and will be removed from the table. The update frequency for the new design point is initially set as 1 in the table to encourage the selection module to select new design points for local search.

## VI. Experiments

### A. Experiment Setup

We compared our proposed MOSAIC framework with three state-of-the-art approaches: simulated annealing-based tri-objective optimization (TOO) [29], genetic algorithm-based load distribution (GALD) [13], and decomposition-based multi-objective evolutionary algorithm with Gaussian mutation and crowding distance (DMGC) [30]. We extended these frameworks to our multi-objective problem to co-optimize energy cost, carbon emissions, and water consumption, by distributing workloads geographically and managing energy payment plans subsequently. We evaluated all frameworks by determining their energy costs, carbon emissions, and water usage, as well as the Pareto Hypervolume (PHV) [28] of solutions.

For MOSAIC we make use of the following parameter values in all experiments: $N = 30$, $gen = \infty$, $f_{update} = 50$, and $iter_{early} = 500$. In each datacenter, we consider three different types of Intel server nodes: E3-1225v3, E5649, and E5-2697v2, which differ in their number of cores, frequency, power profile, and memory. Each datacenter is assumed to have 4320 computing nodes (servers). The mix of server node types differs across datacenter locations. We considered 16 different datacenter locations across the United States, with diverse energy and cost characteristics, as shown in Fig. 4.

Due to the popularity of data analytics workloads among cloud service providers, we chose five data-intensive workloads from the BigDataBench 5.0 [27]: Latent Dirichlet Allocation (LDA), K-means, Naive Bayes, Image-to-Text, and Image-to-Image. For our power model, we use CoP data from [37] which varies between 3.74 and 5.73. For our water model, 0~3.97 L/kWh $EWIF$ from [43] is adopted. We consider a temperature of 40 Celsius at the water-cooling tower and the corresponding water latent heat at this temperature is 0.66 kWh/L. In the water-cooling tower, the concentration cycle of remaining water is assumed to be 5. The energy intensity in potable water generation and wastewater treatment is estimated to be 550 kWh/ML and 640 kWh/ML respectively according to [52]. We determined pricing data from local power companies' websites across the United States. Based on our analysis, the TOU price varies from 1.8~48¢/kWh, clean premium varies from 0.39~144¢/kWh and the annual clean contract cost is 15¢/kWh.

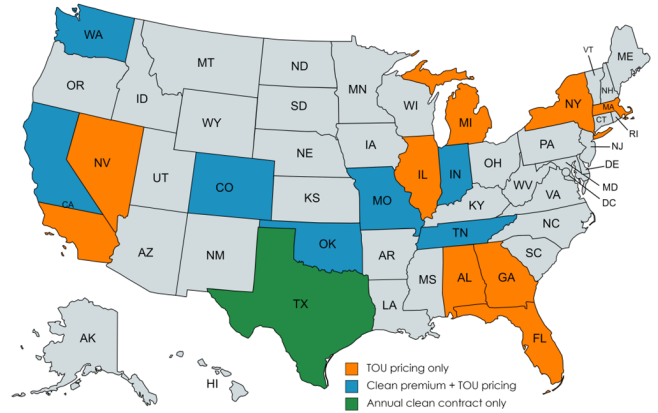

Fig. 4. Datacenter power price map with different price models: TOU price, clean premium, and annual clean contract. States with blue/orange/green color host datacenters that we consider in our studies. Locations in orange are ones with only TOU pricing available. Locations in blue are ones with clean premium projects as well as TOU pricing available. Location in green is the one with annual clean contract available. Locations in gray are not considered in our experiments.

### B. Experimental Results

#### 1) PHV Improvements

To compare the solution quality in multi-objective optimization, we first determined the PHV over time for a scenario with 16 datacenters and all three objectives (minimizing energy cost, carbon emissions, and water usage). PHV is a metric that calculates the hyperspace volume which is enclosed by all solutions on the Pareto front and a user-defined reference point. A higher value of PHV is indicative of a more diverse and higher-quality solution set.

Fig. 5 shows the PHV for MOSAIC and the three comparison frameworks. We can observe that all frameworks converge within ~8 minutes. MOSAIC has a PHV that is 1.53× larger than that of the second-best framework (GALD). Further, to reach the second-best PHV result, it takes MOSAIC 27.45× less time than GALD. Based on these results, we can see that MOSAIC optimizes the PHV better and faster than other design space exploration frameworks.

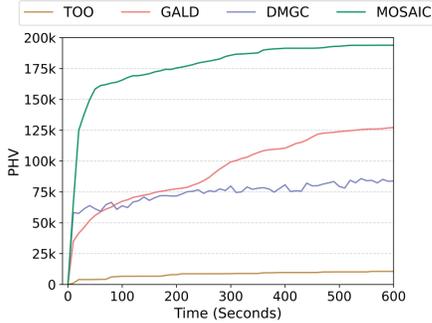

Fig. 5. Three objective (energy cost, carbon emissions, water usage) PHV over time of MOSAIC, GALD, DMGC, and TOO.

To further analyze the differences within PHVs across frameworks, we plot the corresponding 2D Pareto fronts of all frameworks in Fig. 6. In Fig. 6(a), GALD and MOSAIC can be observed to have more diverse Pareto fronts, for normalized energy cost and normalized carbon emission. But MOSAIC's Pareto front dominates GALD's. A similar trend can be observed in Fig. 6(b) where GALD and MOSAIC have a more diverse Pareto front in terms of normalized energy cost and normalized water usage. But MOSAIC's Pareto front is far more dominant than GALD's front. From these plots, we can summarize that MOSAIC not only generates a more diverse Pareto optimal solution set but also dominates Pareto fronts of other frameworks.

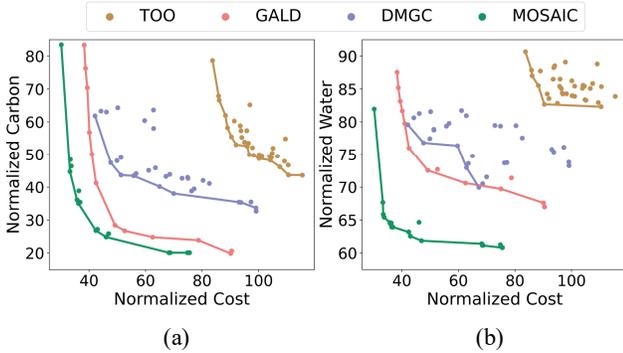

Fig. 6. Pareto front comparison of frameworks for: (a) carbon emissions vs. energy cost; (b) water use vs energy cost

*2) Scalability Analysis*

To evaluate each framework's performance scalability across different design spaces and workload assumptions, we further analyzed the frameworks across multiple scenarios. We changed the size and complexity of the design space by modifying the number of datacenters (Fig. 7(a)) and subscription rate of datacenters (Fig. 7(b)). The average PHV of the TOO framework is used as the baseline, and we normalize the PHVs of other frameworks to it.

In Fig. 7(a), we compare the PHV for three scenarios with 4, 8, and 16 datacenters considered in our optimization problem. MOSAIC outperforms other frameworks and interestingly, does not experience the same performance degradation as other frameworks do in the 16 datacenter scenario. In such a scenario, MOSAIC's PHV is 2.21× higher than the second-best framework (DMGC). This shows that MOSAIC has good scalability with increasing datacenter size.

From Fig. 7(b), we can observe a similar phenomenon. As the workload subscription rate changes from low (50%) to medium (75%), and high (99%), MOSAIC shows the best PHV improvement, and the differences between MOSAIC and other frameworks enlarge with the subscription rate. For the high subscription rate, MOSAIC's PHV is 2.20× higher than the second-best framework (DMGC).

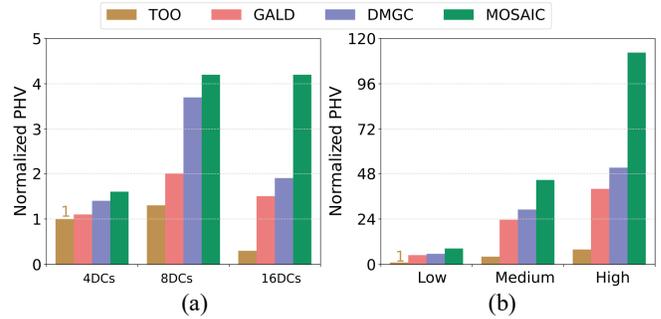

Fig. 7. Sensitivity analysis on: (a) number of datacenters and (b) workload subscription rate of datacenters. All PHV values are normalized w.r.t. the TOO framework.

After evaluating the frameworks' scalability in terms of design space, we further examine their scalability in terms of the number of optimization objectives. We change the number of objectives from one (1obj; energy cost) to two (2obj; energy cost and carbon emissions), and three (3obj; energy cost, carbon emissions, water use). Fig. 8 shows the results for these three scenarios across all frameworks, with results normalized to the PHV for the TOO framework. It can be observed that MOSAIC again outperforms other frameworks and has good scalability with the number of objectives. In the 3-obj scenario, MOSAIC's PHV is 2.20× higher than the second-best framework (DMGC).

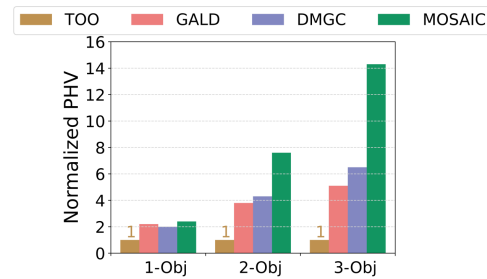

Fig. 8. Sensitivity analysis on number of optimization objectives. All PHV values are normalized w.r.t. the TOO framework.

*3) Solution Quality Analysis*

While PHV provides a proxy estimate of the quality of solutions generated by a framework, it is important to also analyze the actual solutions generated by the framework, across metrics of interest. In our next experiment, we compared the performance of all frameworks across the three objective values of energy cost, water use, and carbon emissions, for a problem scenario with 16 datacenters and 10-minute runtime constraint for all frameworks in each epoch.

Table I shows results aggregated over a 24-hour period, with values normalized to the TOO framework. The results are categorized into three sets: cost-efficient (the solution in the output Pareto set of each framework with the lowest energy cost), carbon-efficient (the solution in the output Pareto set of each framework with the lowest carbon emissions), and water-efficient (the solution in the output Pareto set of each framework with the lowest water use).

TABLE I. 24-HOUR AGGREGATED RESULTS FOR ENERGY COST-, CARBON-, AND WATER-EFFICIENT SOLUTIONS

|  | Cost-Efficient | | | Carbon-Efficient | | | Water-Efficient | | |
| --- | --- | --- | --- | --- | --- | --- | --- | --- | --- |
|  | Cost | Carbon | Water | Cost | Carbon | Water | Cost | Carbon | Water |
| **TOO** | 1.00 | 1.00 | 1.00 | 1.15 | 0.57 | 0.93 | 1.11 | 0.68 | 0.89 |
| **GALD** | 0.89 | 0.68 | 0.85 | 0.91 | 0.58 | 0.82 | 0.91 | 0.59 | 0.82 |
| **DMGC** | 0.79 | 0.86 | 0.91 | 1.01 | 0.56 | 0.81 | 0.98 | 0.61 | 0.77 |
| **MOSAIC** | 0.75 | 1.23 | 1.01 | 0.90 | 0.32 | 0.73 | 0.91 | 0.36 | 0.71 |

We also determine a single best solution from the Pareto set generated by each framework that has the lowest cumulative energy cost, carbon emission, and water usage for each epoch. Each framework's single best solutions are aggregated, and we calculate the overall cumulative results for a 24-hour period. We normalize the overall results with respect to the TOO framework as shown in Fig. 9. Over a 24-hour interval, we find that MOSAIC improves cumulative solution quality by 4.61×, 2.20×, and 3.16× when compared to TOO, GALD, and DMGC, respectively. MOSAIC thus has the best optimization performance for this three-objective problem compared to state-of-the-art frameworks.

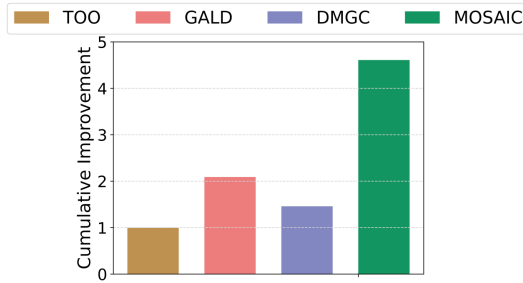

Fig. 9. 24-hour cumulative improvement of three objectives (energy cost, carbon emission, and water usage) for the best solution generated by a framework w.r.t. the TOO framework.

## VII. CONCLUSION

In this work, for the first time, we have studied the multi-objective sustainable datacenter management problem to simultaneously minimize energy cost, carbon footprint, and water use of datacenters. Comprehensive models of energy consumption, energy price, water consumption, carbon emissions, and workload execution were built to support realistic sustainable datacenter management. We then developed a novel framework called MOSAIC to manage workload distribution and datacenter operation. In experiments, MOSAIC was able to provide better quality solutions and arrive at these solutions more quickly than other frameworks. MOSAIC was able to realize 27.45× speedup and 1.53× improvement in PHV while reducing the carbon footprint by up to 3.09×, water footprint by up to 1.40×, energy costs by up to 1.33×. and a cumulative improvement across all objectives (carbon, water, cost) of up to 4.61×, compared to state-of-the-art frameworks.

ACKNOWLEDGEMENT

This research was made possible with support from HPE and grants from the National Science Foundation (CCF-2324514, CNS-2132385). We would also like to acknowledge Ninad Hogade for his help with the evaluation framework.